\documentclass[journal]{IEEEtran}
\usepackage{amssymb}

\usepackage{subfigure}
\usepackage{graphicx}
\usepackage{enumerate}
\usepackage{amsmath}
\usepackage{amsfonts}
\usepackage{algorithm}
\usepackage{algpseudocode}
\usepackage{url}
\usepackage{epstopdf}
\usepackage{multirow}
\usepackage{authblk}
\usepackage{subfigure}
\usepackage{xcolor}
\usepackage{mathrsfs}

\usepackage{verbatim}

\newtheorem{theorem}{Theorem}[section]
\newtheorem{example}[theorem]{Example}
\newtheorem{corollary}[theorem]{Corollary}
\newtheorem{proposition}[theorem]{Proposition}
\newtheorem{definition}[theorem]{Definition}
\newtheorem{lemma}[theorem]{Lemma}
\newcommand{\prof}{\begin{IEEEproof}}
\newcommand{\eprof}{\end{IEEEproof}}

\newcommand{\prop}{\begin{proposition}}
\newcommand{\eprop}{\end{proposition}}
\newcommand{\them}{\begin{theorem}}
\newcommand{\ethem}{\end{theorem}}
\newcommand{\dfn}{\begin{definition}}
\newcommand{\edfn}{\end{definition}}
\newcommand{\exm}{\begin{example}}
\newcommand{\eexm}{\end{example}}
\newcommand{\coro}{\begin{corollary}}
\newcommand{\ecoro}{\end{corollary}}
\newcommand{\lem}{\begin{lemma}}
\newcommand{\elem}{\end{lemma}}

\newcommand{\eps}{\varepsilon}

\ifCLASSINFOpdf
\else
\fi
\begin{document}
%

\title{Verification of infinite-step and K-step opacity Using Petri Nets}
%
%
%

\author{Hao~Lan, 
        Yin~Tong, ~\IEEEmembership{Member,~IEEE}
        ~Jin~Guo
        and Carla~Seatzu, \IEEEmembership{Senior Member,~IEEE}%
\thanks{H. Lan, Y. Tong (Corresponding Author), and Jin Guo are with the School of Information Science and Technology, Southwest Jiaotong University, Chengdu 611756, China
        {\tt\small haolan@my.swjtu.edu.cn; yintong@swjtu.edu.cn; jguo\_scce@swjtu.edu.cn}}
\thanks{C. Seaztu is with the Department of Electrical and Electronic Engineering,
University of Cagliari, 09123 Cagliari, Italy {\tt\small seatzu@diee.unica.it}}}%

%
%

\markboth{
}%
{Shell \MakeLowercase{\textit{et al.}}: Bare Demo of IEEEtran.cls for IEEE Journals}
%



\maketitle

\begin{abstract}
This paper addresses the problem of infinite-step opacity and K-step opacity of discrete event systems  modeled with Petri nets. A Petri net system is said to be infinite-step/K-step opaque if all its secret states remains opaque to an intruder for any instant within infinite/K steps. In other words, the intruder is never able to ascertain that the system used to be in a secrete state within infinite/K steps based on its observation of the system¡¯s evolution. Based on the notion of basis reachability and the two-way observer, an efficient approach to verify infinite-step opacity and K-step opacity is proposed.
\end{abstract}


%
\IEEEpeerreviewmaketitle

\section{Introduction}

Motivated by the concern about security and privacy, opacity has been wildly investigated in the past years \cite{bryans2005modelling,lin2011opacity,wu2013comparative,saboori2013initial,jacob2015opacity}. Opacity describes the ability of a system to hide its secret behavior from the intruders. Different notions of opacity properties have been defined for discreat event systems (DESs), including
language-based opacity \cite{lin2011opacity,zhang2012polynomial},
current-state opacity \cite{bryans2005modelling,wu2013comparative},
initial-state opacity \cite{saboori2013initial,tong2015initial},
K-step opacity \cite{falcone2015enforcement,yin2019infinite},
infinite-step opacity \cite{saboori2011infinite,yin2017new}, etc.
In particular, we discuss \emph{K-step opacity} and \emph{infinite-step opacity} here.

Given a set of secret states, based on the observation of a system¡¯s evolution, if the system can not be inferred that it used to reach one of the secret states at any moment, the system is \emph{infinite-step opaque}. Analogously, a system is \emph{K-state opaque} if, given a set of secret states, by observing the sequence of events generated by the system, the intruder will not be able to infer that the system used to reach one of the secret states within K steps.

The notion of K-step opacity was first proposed in \cite{saboori2007k} in the nondeterministic finite automaton framework based on the assumption that the events are partially observable. Then Saboori and Hadjicostis \cite{saboori2009in} characterized the notion of infinite-step opacity as an extension of the notion of K-step opacity. Later, they explore the two opacity properties deeply in \cite{saboori2011infinite,saboori2011K}.
Saboori and Hadjicostis \cite{saboori2011infinite} have shown that infinite-step opacity can be verified by constructing a current-state estimator and a bank of initial-state estimators for a given nondeterministic finite automaton, and the verification of infinite-step opacity is proved to be PSPACE-hard. In \cite{saboori2011K}, K-delay state estimator of the system is introduced to check K-step opacity with complexity of ${\cal O}((|E_o|+1)^K \times |E_o| \times 2^{|X|})$, where $X$ is the set of states and $E_o$ is the set of observable events of the system.
Furthermore, more efficient approaches are proposed to check both infinite-step opacity and K-step opacity in \cite{yin2017new,yin2016two}. The approaches are based on the construction of a new tool, called ¡°two-way observer¡±. The two-way observer (TW-observer) is built by concurrent composition of two observers, one is the observer of the given automation, another is the observer of the reverse automaton. Yin and Lafortune \cite{yin2017new} show that infinite-step opacity can be verified with complexity of ${\cal O}(|E_o| \times 2^{|X|} \times 2^{|X|})$ and K-step opacity can be verified with complexity of ${\cal O}(min\{2^{|X|},|E_o|^K\} \times |E_o| \times 2^{|X|})$. The notation of two opacity properties is also extended to stochastic DESs \cite{yin2019infinite}, and the enforcement of the K-step opacity is proposed in \cite{falcone2015enforcement}.

Petri nets have been wildly used to model and check different types of opacity, e.g., initial-state opacity \cite{tong2015initial}, current state opacity \cite{tong2015current}, and language-based opacity \cite{tong2016verification}. Using structural analysis and algebraic techniques, these problems can be solved more efficiently by Petri net and its basis reachability graph (BRG).
To the best of our knowledge, currently there is no work that study infinite-step opacity and K-step opacity in labeled Petri nets (LPNs).

In this paper, the formalization and verification of infinite-step opacity and K-step opacity in bounded labeled Petri nets are addressed. The secret is defined as a subset of the reachable markings. A labeled Petri net is infinite/K-step opacity opaque with respect to a secret if the intruder can never infer that the observed sequence used to origins from a secret marking within infinite/K steps.
Considering that a possible non-secret marking that is reachable from a secret basis marking by firing only unobservable transitions is unable to be distinguished by the intruder, we make the following reasonable assumption: if a basis marking belongs to a secret, then all markings in its unobservable reach belong to the secret. Then we prove that infinite-step opacity and K-step opacity can be checked by using the BRG of the system. We present necessary and sufficient conditions for infinite-step opacity and K-step opacity, by analyzing the TW-observer of the BRG of the original LPN system. Since BRG is usually much smaller than the reachability grach (RG), this leads to a relevant advantage in terms of computational complexity.
To reduce the complexity, we propose a new structure called modified TW-observer to verify infinite-step opacity.
In the paper, we first extend the two opacity properties to labeled Petri nets and then based on the notion of basis marking efficient approaches to verify the two opacity properties are proposed.
The contributions of the work are summarized as follows.

\begin{itemize}
  \item Infinite-step opacity and K-step opacity are formally defined in labeled Petri net systems.
  \item Under a reasonable assumption, efficient approaches to verify the above two opacity properties in bounded labeled Petri nets are proposed. Based on basis markings, enumerating all the markings that consistent with an observation is avoided. By constructing the TW-observer of the BRG, the two opacity properties can be checked.
  \item Differently from \cite{yin2017new}, we propose a modified TW-observer to check infinite-step opacity, whose computational complexity is lower than the construction method in \cite{yin2017new}.
\end{itemize}

The rest of the paper is organized as follows.
In Section~\ref{sec:pre} background on finite automata and labeled Petri nets are recalled.
Infinite-step opacity and K-step opacity in labeled Petri nets are defined in Section~\ref{sec:def}.
In Section~\ref{sec:brg}, the fomalizition and property of the BRG are presented.
Efficient approaches to verify Infinite-step opacity and K-step opacity are proposed in Section~\ref{sec:ver}. Conclusions are finally drawn in Section~\ref{sec:con} where our future lines of research in this framework are illustrated.

\section{Preliminaries and Background}\label{sec:pre}
In this section we recall the formalisms used in the paper and some results on state estimation in labeled Petri nets. For more details, we refer to \cite{cabasino2011discrete,murata1989petri,cassandras2009introduction}.

\subsection{Automata}\label{subsec:auto}
A \emph{nondeterministic finite (state) automaton} (NFA) is a 4-tuple $A=(X, E_A, f, x_0)$, where $X$ is the finite \emph{set of states}, $E_A$ is the finite \emph{set of events}, $f: X\times E_{\eps}\rightarrow 2^X$ is the (partial) \emph{transition relation}, $E_{\eps}=E_A \cup \{\eps\}$, and $x_0\in X$ is the \emph{initial state}. The transition relation $f$ can be extended to $f:X\times E_{\eps}^*\rightarrow 2^X$ in a standard manner. Given an event sequence $w\in E_{\eps}^*$, if $f(x_0,w)$ is defined in $A$, $f(x_0,w)$ is the set of states reached in $A$ from $x_0$ with $w$ occurring.
We denote as $A_R=(X, E_A, f_R, X)$ the reverse automation of $A$. The reverse automation $A_R$ is constructed by revising all arcs in $A$ without specifying the initial states.

Given an NFA, its equivalent DFA, called \emph{observer}, can be constructed following the procedure in Section~2.3.4 of \cite{cassandras2008introduction}. Each state of the observer is a subset of states of $X$ in which the NFA may be after a certain event sequence has occurred. The complexity of computing the observer is ${\cal O}(2^n)$, where $n$ is the number of states of $A$.

\subsection{Petri Nets}
A \emph{Petri net} is a structure $N=(P,T,Pre,Post)$, where $P$ is a set of $m$ \emph{places}, graphically represented by circles; $T$ is a set of $n$ \emph{transitions}, graphically represented by bars; $Pre:P\times T\rightarrow\mathbb{N}$ and $Post:P\times T\rightarrow\mathbb{N}$ are the \emph{pre-} and \emph{post-incidence functions} that specify the arcs directed from places to transitions, and vice versa. The incidence matrix of a net is denoted by $C=Post-Pre$. A Petri net is \emph{acyclic} if there are no oriented cycles.

A \emph{marking} is a vector $M:P\rightarrow \mathbb{N}$ that assigns to each place a non-negative integer number of tokens, graphically represented by black dots. The marking of place $p$ is denoted by $M(p)$. A marking is also denoted by $M=\sum_{p\in P}M(p)\cdot p$. A \emph{Petri net system} $\langle N,M_0\rangle$ is a net $N$ with \emph{initial marking} $M_0$.

A transition $t$ is \emph{enabled} at marking $M$ if $M\geq Pre(\cdot,t)$ and may fire yielding a new marking $M'=M+C(\cdot,t)$. We write $M[\sigma\rangle$ to denote that the sequence of transitions $\sigma=t_{j1}\cdots t_{jk}$ is enabled at $M$, and $M[\sigma\rangle M'$ to denote that the firing of $\sigma$ yields $M'$. The set of all enabled transition sequences in $N$ from marking $M$ is $L(N,M)=\{\sigma\in T^*| M[\sigma\rangle \}$. Given a transition sequence $\sigma\in T^*$, the function $\pi:T^*\rightarrow \mathbb{N}^n$ associates with $\sigma$ the Parikh vector $y=\pi(\sigma)\in\mathbb{N}^n$, i.e., $y(t)=k$ if transition $t$ appears $k$ times in $\sigma$. Given a sequence of transitions $\sigma\in T^*$, its \emph{prefix}, denoted by $\sigma'\preceq \sigma$, is a string such that $\exists \sigma''\in T^*:\sigma'\sigma''=\sigma$. The \emph{length} of $\sigma$ is denoted by $|\sigma|$.

A marking $M$ is \emph{reachable} in $\langle N,M_0\rangle$ if there exists a transition sequence $\sigma$ such that $M_0[\sigma\rangle M$. The set of all markings reachable from $M_0$ defines the \emph{reachability set} of $\langle N,M_0\rangle$, denoted by $R(N,M_0)$. Given a marking $M\in R(N,M_0)$, we define $$UR(M)=\{M'\in \mathbb{N}^{m} | M[\sigma_u\rangle M',\sigma_u\in T^*_u\}$$ its \emph{unobservable reach}, the set of markings reachable from $M$ through unobservable transitions. A Petri net system is \emph{bounded} if there exists a non-negative integer $k \in \mathbb{N}$ such that for any place $p \in P$ and any reachable marking $M \in R(N,M_0)$, $M(p)\leq k$ holds.

A \emph{labeled Petri net} (LPN) system is a 4-tuple $G=(N,M_0,\allowbreak E,\ell)$, where $\langle N,M_0\rangle$ is a Petri net system, $E$ is the \emph{alphabet} (a set of labels) and $\ell:T\rightarrow E\cup\{\eps\}$ is the \emph{labeling function} that assigns to each transition $t\in T$ either a symbol from $E$ or the empty word $\eps$. Therefore, the set of transitions can be partitioned into two disjoint sets $T=T_o\cup T_u$, where $T_o=\{t\in T|\ell(t)\in E\}$ is the set of observable transitions with $|T_o|=n_o$ and $T_u=T\setminus T_o=\{t\in T|\ell(t)=\eps\}$ is the set of unobservable transitions with $|T_u|=n_u$.
The labeling function can be extended to transition sequences $\ell: T^*\rightarrow E^*$ as $\ell(\sigma t)=\ell(\sigma)\ell(t)$ with $\sigma\in T^*$ and $t\in T$. Given a set $Y\subseteq R(N,M_0)$ of markings, the \emph{language generated by} $G$ \emph{from} $Y$ is
$${\cal L}(G,Y)=\bigcup_{M\in Y}\{w\in E^*| \exists \sigma \in L(N,M): w=\ell(\sigma)\}.$$
In particular, the \emph{language generated by} $G$ is
$${\cal L}(G,\{M_0\})=\{w\in E^*|\exists \sigma\in L(N,M_0):w=\ell(\sigma)\},$$
that is also simply denoted by ${\cal L}(G)$. Let $w\in {\cal L}(G)$ be an observed word. We denote as $${\cal C}(w)=\{M\in \mathbb{N}^m|\exists \sigma\in L(N,M_0):M_0[\sigma\rangle M, \ell(\sigma)=w\}$$  the set of markings \emph{consistent} with $w$. 

Given an LPN system $G=(N,M_0,\allowbreak E,\ell)$ and the set of unobservable transitions $T_u$, the \emph{$T_u$-induced subnet} $N'=(P,T',\allowbreak Pre',Post')$ of $N$, is the net resulting by removing all transitions in $T\setminus T_u$ from $N$, where $Pre'$ and $Post'$ are the restriction of $Pre$, $Post$ to $T_u$, respectively. The incidence matrix of the $T_u$-induced subnet is denoted by $C_u=Post'-Pre'$.

\section{Infinite-step opacity and K-step opacity in Labeled Petri Nets}\label{sec:def}

Infinite-step opacity and K-step opacity have been defined in automation \cite{saboori2011infinite,yin2017new,saboori2011K}. In this section we extend these two opacity properties to labeled Petri nets.


In the framework of LPN system, we denote a \emph{secret} as a set of reachable makings $S \subseteq R(N,M_0)$. A marking $M \in S$ is a \emph{secret marking}. Markings in $\bar{S}=R(N,M_0) \setminus S$ are \emph{non-secret markings}.

\dfn\label{def:ISP}
[\textbf{Infinite-Step Opacity}] Let $G=(N,M_0,\allowbreak E,\ell)$ be an LPN system and $S \subseteq R(N,M_0)$ be a secret. System $G$ is \emph{infinite-step opacity} with respect to $S$ if $\forall \sigma_1\sigma_2 \in L(G)$ with $M_0[\sigma_1\rangle M_1 \in S$, there exists $\sigma_1'\sigma_2' \in L(G)$ such that $M_0[\sigma_1'\rangle M_1' \notin S$,
where $\ell(\sigma_1)=\ell(\sigma_1')$, $\ell(\sigma_2)=\ell(\sigma_2')$.
\hfill $\diamond$
\edfn


%

In words, an LPN system is infinite-step opaque if for any marking $M \in S$ reaching from the initial marking, that there exists a marking $M'$ with the same observation that is not belong to $S$, and $M, M'$ can generate same language.
Namely, the system is infinite-step opacity if the intruder cannot infer that the system used to reach a state that is belong to the secret.

\dfn\label{def:KSP}
[\textbf{K-Step Opacity}] Let $G=(N,M_0,E,\allowbreak\ell)$ be an LPN system, $K\in \mathbb{N}$ be a integer and $S \subseteq R(N,M_0)$ be a secret. System $G$ is \emph{K-step opaque} with respect to $S$ if $\forall \sigma_1\sigma_2 \in L(G)$ with $ M_0[\sigma_1\rangle M_1 \in S$ and $|\ell(\sigma_2)|\leq K$, there exists $\sigma_1'\sigma_2' \in L(G)$ such that
$M_0[\sigma_1'\rangle M_1' \notin S$,
where $\ell(\sigma_1)=\ell(\sigma_1'),~\ell(\sigma_2)=\ell(\sigma_2')$.
\hfill $\diamond$
\edfn


In words, an LPN system is K-step opaque if for any marking $M \in S$ reaching from the initial marking, that there exists a marking $M'$ with the same observation that is not belong to $S$, and any word generated by $M$ within K steps, there are always same word generated by $M'$.
Namely, the system is K-step opaque if the intruder cannot infer that the system used to reach a state that is belong to the secret within K steps. Clearly, when $K=\infty$, it becomes infinite-step opacity, and when $K=0$, it becomes curent-state opacity \cite{tong2015current}.

\begin{figure}
  \centering
  \subfigure[]{%
  \label{fig:LPN}
  \includegraphics[width=0.4\textwidth]{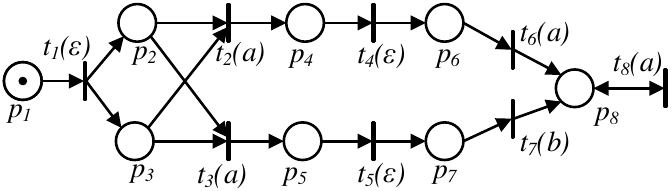}}
  \centering
  \subfigure[]{%
  \label{fig:RG}
  \includegraphics[width=0.47\textwidth]{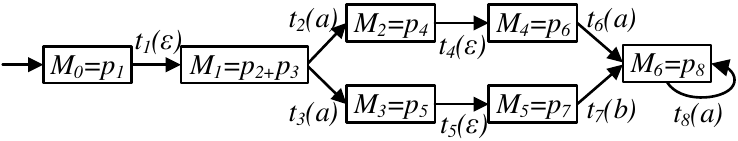}}

  \caption{The LPN system in Example~\ref{eg:opacity} (a), and its RG (b).}
  \label{fig:LL}
\end{figure}

\exm\label{eg:opacity}
Let us consider the LPN system in Fig.~\ref{fig:LPN} where the observable transitions is $T_o=\{t_2, t_3, t_6, t_7, t_8\}$ and the unobservable transitions is $T_u=\{t_1, t_4, t_5\}$. Transitions $t_2$, $t_3$, $t_6$ and $t_8$ are labeled $a$, transition $t_7$ is labeled $b$. The RG of the LPN system is shown in Fig.~\ref{fig:RG}. Let the secret be $S=\{M_2,M_4\}$.
Since $M_0[t_1t_2\rangle M_2 \in S$, clearly there exists a transition sequence $t_1t_3$ that is $M_0[t_1t_3\rangle M_3 \notin S$ and $\ell(t_1t_2) = \ell(t_1t_3)$. However, at $M_2$, transition sequence $t_4t_6$ is the only transtion sequence that enabled, while transiton sequence $t_5t_7$ is the only transiton sequence that can fire at $M_3$. Since $\ell(t_4t_6) \neq \ell(t_5t_7)$ and $|\ell(t_4t_6)|=1$, according to Definition~\ref{def:KSP}, $K=0$, i.e., the system is 0-step opaque (of course, not infinite-step opaque).
\hfill $\diamond$
\eexm

In the following, based on the given secret, we define the secret language and the non-secret language.

We denote as $S(w)={\cal C}(w) \cap S$ the set of secret markings consistent with a given observation $w\in {\cal L}(G)$, and $\bar{S}(w)={\cal C}(w)\setminus S(w)$ be the set of non-secret markings consistent with $w$. The secret language generated by $S(w)$ is defined as
${\cal L}(G,S(w))= \bigcup_{ M\in S(w)} {\cal L}(G,M)$
and the non-secret language is defined as
${\cal L}(G,\bar{S}(w))= \bigcup_{ M\in \bar{S}(w)} {\cal L}(G,M)$.

\lem\label{lem:ISP}
Let $G=(N,M_0,E,\allowbreak\ell)$ be an LPN system and $S \subseteq R(N,M_0)$ be a secret. System $G$ is infinite-step opaque with respect to $S$ if and only if $\forall w\in {\cal L}(G)$, such that ${\cal L}(G,S(w)) \subseteq {\cal L}(G,\bar{S}(w))$.
\elem

\prof
Follows from Definitions~\ref{def:ISP}.
\eprof

In a simple word, an LPN system is infinite-step opaque with respect to a given secret if and only if for any observation, its corresponding secret language is a subset of the non-secret language.

\lem\label{lem:KSP}
Let $G=(N,M_0,E,\allowbreak\ell)$ be an LPN system and $S \subseteq R(N,M_0)$ be a secret. System $G$ is K-step opaque with respect to $S$ if and only if $\forall w\in {\cal L}(G)$, such that $\forall w'\in {\cal L}(G,S(w))$ with $|w'|\leq K$, that $w' \in {\cal L}(G,\bar{S}(w))$.
\elem

\prof
Follows from Definitions~\ref{def:KSP}.
\eprof

In a simple word, an LPN system is K-step opaque with respect to a given secret if and only if for any observation,
any word generated within K steps in its corresponding secret language is also belong to the non-secret language.

Therefore by Lemmas~\ref{lem:ISP} and~\ref{lem:KSP}, the infinite-step opacity and K-step opacity problem in LPN systems is equivalent to the language containment problem.

\section{Basis reachability graph}\label{sec:brg}
In the automaton framework, two-way observer (TW-observer) is used to verify the infinite-step opacity and K-step opacity \cite{yin2017new}. Obviously, in the case of bounded LPN system the same approach can be used first constructing the RG of the net system and then computing its TW-observer.
However, the complexity of constructing the RG of a Petri net system is exponential in the size of the net (number of places, transitions, tokens in the initial marking) and the approach in \cite{yin2017new} has the complexity of ${\cal O}(|E_o| \times 2^{|X|} \times 2^{|X|})$, where $E_o$ is the set of events, and $X$ is the set of states. Thus, such an approach could be unfeasible in the case of systems with a large state space.
In this paper, we propose a new approach based on the notion of basis marking and basis reachability graph to check the above two opacity properties, thus enumerating all states in RG is avoided.

In this section, using the notion of basis marking, we introduce the fomalizition and the property of the BRG for opacity. Then under a reasonable assumption we prove that infinite-step opacity and K-step opacity of the LPN system can be checked by using BRG.
Thus we first review the notion and some results of basis markings, which is proposed in \cite{cabasino2011discrete,ziyue2017basis}.

\dfn\label{def:exp}
Given a marking $M$ and an observable transition $t\in T_o$, we denote as $$\Sigma(M,t)=\{\sigma\in T^*_u|M[\sigma\rangle M',M'\geq Pre(\cdot,t)\}$$ the set of \emph{explanations} of $t$ at $M$ and $Y(M,t)=\{y_u\in \mathbb{N}^{n_u}|\exists \sigma\in \Sigma(M,t):y_u=\pi(\sigma)\}$ the set of \emph{$e$-vectors}. \hfill $\diamond$
\edfn

After firing any unobservable transition sequence in $\Sigma(M,t)$ at $M$, the transition $t$ is enabled.
To provide a compact representation of the reachability set, we are interested in finding the explanations whose firing vector is minimal.

\dfn\label{def:minexp}
Given a marking $M$ and an observable transition $t\in T_o$, we denote as
$$\Sigma_{min}(M,t)=\{\sigma\in \Sigma(M,t)|\nexists \sigma'\in \Sigma(M,t):\pi(\sigma')\lneqq\pi(\sigma)\}$$
the set of \emph{minimal explanations} of $t$ at $M$ and $Y_{min}(M,t)=\{y_u\in \mathbb{N}^{n_u}|\exists \sigma\in \Sigma_{min}(M,t):y_u=\pi(\sigma)\}$ as the corresponding set of \emph{minimal $e$-vectors}. \hfill $\diamond$
\edfn

There are many approaches to calculate $Y_{min}(M,t)$. In particular, Cabasino \emph{et al} present an approach that only requires algebraic manipulations when the $T_u$-induced subnet is acyclic \cite{cabasino2011discrete}.

\dfn\label{def:basisM}
Given an LPN system $G=(N,M_0,E,\ell)$ whose $T_u$-induced subnet is acyclic, its \emph{basis marking set} ${\cal M}_b$ is defined as follows:
\begin{itemize}
  \item $M_0\in {\cal M}_b$;
  \item If $M\in {\cal M}_b$, then $\forall t\in T_o, y_u\in Y_{min}(M,t)$,
  $$M'=M+C(\cdot,t)+C_u\cdot y_u \Rightarrow M'\in {\cal M}_b.$$
\end{itemize}
A marking $M_b\in {\cal M}_b$ is called a \emph{basis marking} of $G$.
\hfill $\diamond$
\edfn

The set of basis markings contains the initial marking and all other markings that are reachable from a basis marking by firing a transition sequence $\sigma_u t$, where $t\in T_o$ is an observable transition and $\pi(\sigma_u)=y_u$ is a minimal explanation of $t$ at $M$. Note that $t$ is enabled at some marking in the unobservable reach of $M$. Clearly, ${\cal M}_b\subseteq R(N,M_0)$, and in practical cases the number of basis markings is much smaller than the number of reachable markings \cite{cabasino2011discrete,ziyue2017basis,tong2017verification}.
And the number of basis markings is finite if the corresponding LPN system is bound.
We denote as ${\cal C}_b(w)={\cal M}_b\cap {\cal C}(w)$ the set of basis markings corresponding to a given observation $w\in {\cal L}(G)$.


To guarantee that the BRG is finite, we assume that the LPN system is bounded.
Based on Definition~\ref{def:basisM}, we denote as $B=(X,E,f,x_0)$ the BRG of a bounded LPN system $G=(N,M_0,E,\ell)$. $X={\cal M}_b$ is a finite set of states, $x_0\in X$ is the initial state of the BRG. The event set of the BRG is the alphabet $E$. The transition function $f: X\times E \rightarrow X$ can be determined by the following rule.
If at marking $M_{b}$ there is an observable transition $t$ for which a minimal explanation exists, then we compute the markings reached firing $t$ and its minimal explanations. Let $M_{b}'$ be one of such markings, then an edge from node $M_{b}$ to node $M_{b}'$ labeled $\ell(t)$ is defined in the BRG. The BRG of the LPN system can be constructed by applying the algorithm in \cite{cassandras2009introduction}.

We denote the language generated by BRG $B$ from a basis marking $M_b$ as ${\cal L}(B,M_b)$. According to the construction of the BRG, if a marking $M \in UR(M_b)$ in $G$, then ${\cal L}(G,M) \subseteq {\cal L}(B,M_b)$.
%
%
%
%
Given a BRG $B=(X,E,f,x_0)$, we denote as $B_R=(X,E,f_r,X)$ the reversed BRG. The initial state of $B_R$ is the entire state space $X$.

\begin{figure}
  \centering
  \subfigure[]{%
  \label{fig:BRG}
  \includegraphics[width=0.33\textwidth]{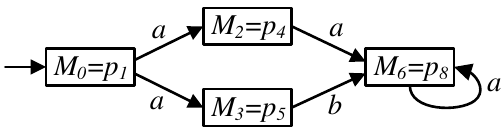}}
  \centering
  \subfigure[]{%
  \label{fig:RB}
  \includegraphics[width=0.3\textwidth]{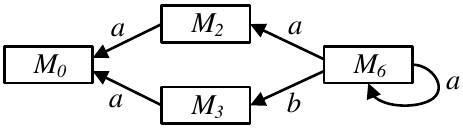}}

  \caption{BRG of the LPN system Fig.~\ref{fig:LL} (a), and the reversed BRG (b).}
  \label{fig:BB}
\end{figure}

\exm\label{eg:BRG}
Let us consider again the LPN system in Fig.~\ref{fig:LPN} whose $T_u$-induced subnet is acyclic, the RG of the LPN system is shown in Fig.~\ref{fig:RG}. The LPN system has 7 reachable markings and only 4 of them are basis markings ${\cal M}_b=\{M_0,M_2,M_3,M_6\}$. The corresponding BRG is presented in Fig.~\ref{fig:BRG}, and the reversed BRG is shown in Fig.~\ref{fig:RB}.
\hfill $\diamond$
\eexm

\dfn\label{def:Sbw}
Let $G=(N,M_0,E,\allowbreak\ell)$ be an LPN system, ${\cal M}_b$ be the set of basis markings, and $S(w)$ be the set of secret markings consistent with $w$. The \emph{secret basis marking set consistent with $w$} $S_b(w)$ is defined as $S_b(w)=S(w) \cap {\cal M}_b$, and the \emph{non-secret basis marking set consistent with $w$} $\bar{S}_b(w)$ is defined as $\bar{S}_b(w)=\bar{S}(w) \cap {\cal M}_b$,
\hfill $\diamond$
\edfn

Since $S_b(w) \subseteq S(w)$ and $\bar{S}_b(w) \subseteq \bar{S}(w)$, we have
${\cal L}(B,S_b(w)) \subseteq {\cal L}(G,S(w))$ and ${\cal L}(B,\bar{S}_b(w)) \subseteq {\cal L}(G,\bar{S}(w))$.
However, ${\cal L}(B,S_b(w)) \subseteq {\cal L}(B,\bar{S}_b(w))$ does not imply that ${\cal L}(G,S(w)) \subseteq {\cal L}(G,\bar{S}(w))$. Thus, to use the BRG, we make the following assumption:
$$\text{A1: }\forall M_b\in S, UR(M_b)\subseteq S.$$
In other words, if a basis marking is a secret marking, then the set of the markings in its unobservable reach belong to the secret. Namely, for all secret basis markings there does not exist an unobservable truansition that leads to a non-secret marking.

\prop\label{prop:RB}
Let $G$ be an LPN system whose $T_u$-induced subnet is acyclic, and $S$ be a secret which satisfy Assumption A1. Let $B$ be the BRG and ${\cal M}_b$ be the set of basis markings of $G$. It holds that ${\cal L}(B,\bar{S}_b(w))={\cal L}(G,\bar{S}(w))$.
\eprop
\prof
First, we prove ${\cal L}(B,\bar{S}_b(w)) \subseteq {\cal L}(G,\bar{S}(w))$.
Since $\bar{S}_b(w) \subseteq \bar{S}(w)$, ${\cal L}(B,\bar{S}_b(w)) = {\cal L}(G,\bar{S}_b(w)) \subseteq {\cal L}(G,\bar{S}(w))$.
Now, we prove ${\cal L}(B,\bar{S}_b(w)) \supseteq {\cal L}(G,\bar{S}(w))$.
Let a marking $M \in \bar{S}(w)$, (case 1) if $M\in {\cal M}_b$, then ${\cal L}(G,M)= {\cal L}(B,M)$;
(case 2) if $M\notin {\cal M}_b$, let $M_b$ be the corresponding basis marking of $M$, namely, $M\in UR(M_b)$. If $M_b \in S_b(w)$, by Assumption A1, $M\in S$, thus it is contradicted. Therfore, $M_b \in \bar{S}_b(w)$, since $M\in UR(M_b)$, ${\cal L}(G,M)\subseteq {\cal L}(B,M)$. Therefore, ${\cal L}(G,\bar{S}(w)) \subseteq {\cal L}(B,\bar{S}_b(w))$.
\eprof

In a simple wrod, given an LPN system, if $\bar{S}_b(w)$ is a non-secret basis marking set of $\bar{S}(w)$, the language generated from $\bar{S}(w)$ in RG is equel to the language generated from $\bar{S}_b(w)$ in the corrsponding BRG.

Now accroding to the assumption A1, we can propose the following proposition.

\prop\label{prop:RtoB}
Let $G$ be an LPN system whose $T_u$-induced subnet is acyclic, and $S$ be a secret which satisfy Assumption A1. Let $B$ be the BRG and ${\cal M}_b$ be the set of basis markings of $G$. We have ${\cal L}(G,S(w)) \subseteq {\cal L}(G,\bar{S}(w))$ if and only if ${\cal L}(B,S_b(w)) \subseteq {\cal L}(B,\bar{S}_b(w))$.
\eprop
\prof
(If)
Since ${\cal L}(B,S_b(w))={\cal L}(G,S_b(w))$ and by Proposition~\ref{prop:RB}, ${\cal L}(B,S_b(w)) \subseteq {\cal L}(B,\bar{S}_b(w))$ $\Leftrightarrow$ ${\cal L}(G,S_b(w)) \subseteq {\cal L}(G,\bar{S}(w))$. Let a marking $M \in S(w)$, (case 1) if $M\in {\cal M}_b$, then $M\in S_b(w)$. Thus ${\cal L}(G,M)\subseteq {\cal L}(G,S_b(w))\subseteq {\cal L}(G,\bar{S}(w))$. (Case 2) If $M\notin {\cal M}_b$, let $M_b$ be the corresponding basis marking of $M$, namely, $M\in UR(M_b)$. Thus ${\cal L}(G,M)\subseteq {\cal L}(G,M_b)$. If $M_b\in S$, then $M_b\in S_b(w)$, thus ${\cal L}(G,M)\subseteq {\cal L}(G,M_b)\subseteq {\cal L}(G,S_b(w))\subseteq {\cal L}(G,\bar{S}(w))$; if $M_b\notin S$, then $M_b\in \bar{S}_b(w)$, thus ${\cal L}(G,M)\subseteq {\cal L}(G,M_b)\subseteq {\cal L}(G,\bar{S}(w))$. Therefore ${\cal L}(G,S(w)) \subseteq {\cal L}(G,\bar{S}(w))$.

(Only if)
Since $S_b(w) \subseteq S(w)$, ${\cal L}(B,S_b(w))={\cal L}(G,S_b(w)) \subseteq {\cal L}(G,S(w))$. By assumption ${\cal L}(G,S(w)) \subseteq {\cal L}(G,\bar{S}(w))$, thus ${\cal L}(B,S_b(w)) \subseteq {\cal L}(G,\bar{S}(w))$. According to Proposition~\ref{prop:RB}, ${\cal L}(B,\bar{S}_b(w))={\cal L}(G,\bar{S}(w))$, therefore ${\cal L}(B,S_b(w)) \subseteq {\cal L}(B,\bar{S}_b(w))$
\eprof

In words, by assumption A1, the language containment problem in the RG can be transformed into that in the BRG. Thus we can rewriting the Lemmas~\ref{lem:ISP} and \ref{lem:KSP} to the following two propositons respectively.

\prop\label{prop:bISP}
Let $G=(N,M_0,E,\allowbreak\ell)$ be an LPN system whose $T_u$-induced subnet is acyclic, and $S$ be a secret which satisfy Assumption A1. System $G$ is infinite-step opaque with respect to $S$ if $\forall w\in {\cal L}(G)$ with $S_b(w) \neq \emptyset$, such that ${\cal L}(B,S_b(w)) \subseteq {\cal L}(B,\bar{S}_b(w))$.
\eprop

\prof
Follows from Lemma~\ref{lem:ISP} and Proposition~\ref{prop:RtoB}.
\eprof

\prop\label{prop:bKSP}
Let $G=(N,M_0,E,\allowbreak\ell)$ be an LPN system whose $T_u$-induced subnet is acyclic, and $S$ be a secret which satisfy Assumption A1. System $G$ is K-step opaque with respect to $S$ if $\forall w\in {\cal L}(G)$ with $S_b(w) \neq \emptyset$, such that $\forall w'\in {\cal L}(B,S_b(w))$ with $|w'|\leq K$, that $w' \in {\cal L}(B,\bar{S}_b(w))$.
\eprop

\prof
Follows from Lemma~\ref{lem:KSP} and Proposition~\ref{prop:RtoB}.
\eprof


In other words, Propositions~\ref{prop:bISP} to~\ref{prop:bKSP} proves that the infinite-step opacity and K-step opacity problem in the LPN system is equivalent to the language containment problem in the corresponding BRG. Thus, in the following, we can check the two opacity properties by the analysis of the BRG of the LPN system.

\section{Verification of the two opacity properties}\label{sec:ver}

In this section we first briefly recall a technique that is used to verify infinite-step opacity and K-step opacity in automata \cite{yin2017new}. Based on the result in the previous section, we show that by applying the technique to the BRG of an LPN system, the two opacity properties of the LPN system can be effectively verified.

In \cite{yin2017new} an automaton called two-way observer (TW-observer) is proposed based on the two observers, one is the observer of the original discrete event system and another is the observer of the reverse automaton of the original system (the second observer is also called initial state estimator in \cite{wu2013comparative,tong2015initial}).

We denote as ${\cal B}_o=({\cal{X}},E,f_o,\hat{X}_0)$ the observer of the BRG $B = (X, E, f, x_0)$. The initial-state estimator of the BRG is denoted by ${\cal B}_e=({\cal{X}}_e,E,f_e,\bar{X}_0)$, as mentioned above, the initial-state estimator ${\cal B}_e$ is the observer of the reversed BRG $B_R$.

\begin{figure}
  \centering
  \subfigure[]{%
  \label{fig:Bo}
  \includegraphics[width=0.15\textwidth]{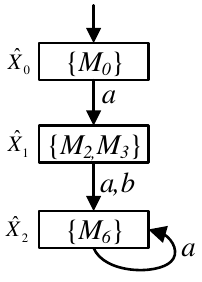}}
  \centering
  \subfigure[]{%
  \label{fig:Be}
  \includegraphics[width=0.3\textwidth]{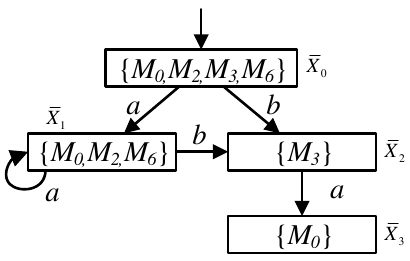}}

  \caption{The observer of the BRG in Fig.~\ref{fig:BB} (a), and the initial-estimator of the BRG (b).}
  \label{fig:ob}
\end{figure}

\exm\label{eg:ob}
Consider again the LPN system in Fig.~\ref{fig:LPN} whose $T_u$-induced subnet is acyclic, the observer of its BRG is presented in Fig.~\ref{fig:Bo}, and the observer of the reversed BRG, i.e., the initial-state estimator is shown in Fig.~\ref{fig:Be}. Given an observation $w = ba$, in the estimator, the reached state is $\{M_0\}$, which implies that the set of states that can generate $w' = ab$ in observer is state $\{M_0\}$ in Fig.~\ref{fig:Bo}.
\hfill $\diamond$
\eexm

\prop\label{prop:obs}
Let $G$ be an LPN system whose $T_u$-induced subnet is acyclic, ${\cal B}_o=({\cal{X}},E,f_o,\hat{X}_0) $ be the observer of its BRG, ${\cal B}_e=({\cal{X}}_e,E,f_e,\bar{X}_0)$ be the initial-state estimator of the BRG, and $S$ be a secret. System $G$ is infinite-step opaque with respect to $S$ if and only if
$$\forall w_1w_2 \in {\cal L}(G): f_o(\hat{X}_0, w_1) \cap f_e(\bar{X}_0, w_2^r)  \nsubseteq S,$$
where $w_2^r$ is the reversed word of $w_2$.
\eprop
\prof
Follow from Proposition 1 and Theorem 2 in \cite{yin2017new}.
\eprof

The above proposition implies that, an LPN system is infinite-step opaque with respect to $S$ if and only if for any intersection of the observer and the initial-state estimator that it is not belong to the secret.

\subsection{Verification of the infinite-step opacity}

In \cite{yin2017new}, infinite-step opacity can be checked by the approach based on the TW-observer. Obviously, the same approach can be used in BRG. However, there are too many transitions in the TW-observer, and we find there is no need for so many transitions to check infinite-step opacity. Thus, we propose an algorithm to build a modified TW-observer which reduces the number of transitions to reduce the complexity.


Given a BRG $B = (X, E, f, x_0)$, the observer of the BRG is ${\cal B}_o=({\cal{X}},E,f_o,\hat{X}_0)$ and the initial-state estimator of the BRG is ${\cal B}_e=({\cal{X}}_e,E,f_e,\bar{X}_0)$.
We denote as ${\cal B}_{tw}=(Q,E_{tw},f_{tw},q_0)$ the modified TW-observer of the BRG. $Q \subseteq {\cal{X}} \times {\cal{X}}_e$ is a finite set of states, the initial state of the modefied TW-observer is the combination of the initial markings of ${\cal B}_o$ and ${\cal B}_e$, that is $q_0=(\hat{X}_0, \bar{X}_0) \in Q$, and each state $q$ in ${\cal B}_{tw}$ consists of two components, we denote as $q=(q(1),q(2)) \in Q$ with the first component $q(1) \in {\cal{X}}$ and the second component $q(2) \in {\cal{X}}_e$.
$E_{tw}=(E \times \{\lambda\})\cup(\{\lambda\} \times E)$ is the event set of the modefied TW-observer. The transition function $f_{tw}: Q \times E_{tw} \rightarrow Q$.

The procedure to construct the modified TW-observer for the opacity is summarized in Algorithm~\ref{alo:TW}, which works as follows.
First, we search the transitions from the initial state $q_0=(\hat{X}_0, \bar{X}_0)$ for the second element.
We search for all the reachable state from the $\bar{X}_0$ to build new states and keep the first component of each new states at the initial state $\hat{X}_0$. And we label each transitions in the form of $(\lambda,e)$ (Steps 3 to 12).
Then we search the transitions from the set of states that have been build, and keep the second element not changed. These transitions are labeled in the form of $(e,\lambda)$ (Steps 13 to 23).
Clearly, the modified TW-observer is a sub-automaton of the TW-observer in \cite{yin2017new}.

\begin{figure}
  \centering
  \includegraphics[width=0.45\textwidth]{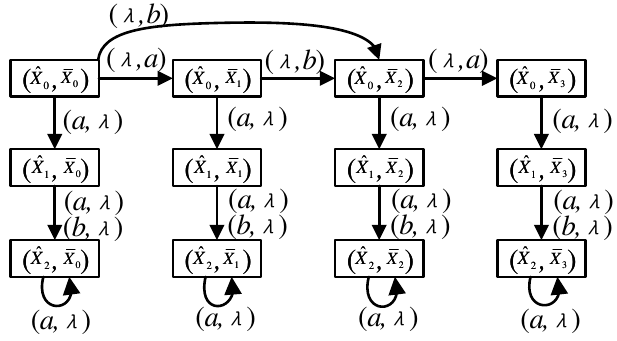}\\
  \caption{The modefied TW-observer of the LPN system in Fig.~\ref{fig:LPN}.}\label{fig:tw}
\end{figure}

\exm\label{eg:tw}
Consider again the LPN system in Fig.~\ref{fig:LPN} whose $T_u$-induced subnet is acyclic. The modefied TW-observer of the LPN system is shown in Fig.~\ref{fig:tw}. In Fig.~\ref{fig:tw}, for example, state $(\hat{X}_2,\bar{X}_1)$ represent state $(\{M_6\},\{M_0,M_2,M_6\})$, which can be reached by string $(\lambda, a)(a, \lambda)(b, \lambda)$.
\hfill $\diamond$
\eexm

\prop\label{prop:tw}
Let $G$ be an LPN system whose $T_u$-induced subnet is acyclic, ${\cal B}_{tw}=(Q,E_{tw},\allowbreak f_{tw},q_0)$ be the modified TW-observer of its BRG, and $S$ be a secret. There exists a state $q=(q(1),q(2))\in Q$ if and only if there exist a state $q(1) \in {\cal{X}}$ and a state $q(2) \in {\cal{X}}_e$.
\eprop
\prof
Follow from the Algorithm~\ref{alo:TW}.
\eprof

\begin{algorithm}
\caption{Computation of the modified TW-observer}
\label{alo:TW}
  \begin{algorithmic}[1]
   \Require
    A observer ${\cal B}_o=({\cal{X}},E,f_o,\hat{X}_0)$;
    A initial-state estimator ${\cal B}_e=({\cal{X}}_e,E,f_e,\bar{X}_0)$.
   \Ensure
    A modified TW-observer ${\cal B}_{tw}=(Q,E_{tw},\allowbreak f_{tw},q_0)$.
   \State $q_0:=(\hat{X}_0,\bar{X}_0)$.
   \State $Q:=\{q_0\}$, $Q_{new}:=\{q_0\}$.
   \ForAll{$q=(q(1),q(2))\in Q_{new}$,}
     \ForAll{$e\in E$: $f_e(q(2),e)!$,}
       \State $q':=(q(1),f_e(q(2),e))$,
       \State $f_{tw}(q,(\lambda,e)):=q'$,
       \If {$q'\notin Q$,}
         \State $Q:=Q \cup \{q'\}$, $Q_{new}:=Q_{new} \cup \{q'\}$,
       \EndIf
     \EndFor
     \State $Q_{new}:=Q_{new} \setminus \{q\}$.
   \EndFor

   \State $Q_{tem}:=Q$.
   \ForAll{$q=(q(1),q(2))\in Q_{tem}$,}
     \ForAll{$e\in E$: $f_o(q(1),e)!$,}
       \State $q':=(f_o(q(1),e),q(2))$,
       \State $f_{tw}(q,(e,\lambda)):=q'$,
       \If {$q'\notin Q$,}
         \State $Q:=Q \cup \{q'\}$, $Q_{tem}:=Q_{tem} \cup \{q'\}$,
       \EndIf
     \EndFor
     \State $Q_{tem}:=Q_{tem} \setminus \{q\}$.
   \EndFor
  \end{algorithmic}
\end{algorithm}

In other words, for an LPN system, there exists a state $q$ in the modefied TW-observer of its BRG if and only if the first element of $q$ exists in the observer, while the second element of $q$ exists in the initial-state estimator.


\them\label{therm:ISO}
Let $G$ be an LPN system whose $T_u$-induced subnet is acyclic, ${\cal B}_{tw}=(Q,E_{tw},\allowbreak f_{tw},q_0)$ be the modefied TW-observer of its BRG, and $S$ be a secret. System $G$ is infinite-step opaque with respect to $S$ if and only if $\forall q=(q(1),q(2))\in Q$, such that
$$q(1)\cap q(2) \nsubseteq S \vee q(1)\cap q(2) = \emptyset$$
\ethem
\prof
Follow from the Propositions~\ref{prop:obs} and~\ref{prop:tw}.
\eprof

In simple words, an LPN system is infinite-step opaque with respect to $S$ if and only if for any state $q$ in the modefied TW-observer such that the intersection of the first and second elements of $q$ does not belong to the secret or is empty.
\exm\label{eg:ISO}
Consider again the LPN system in Fig.~\ref{fig:LPN} whose $T_u$-induced subnet is acyclic, where the secret $S=\{M_2,M_4\}$. The modefied TW-observer of the LPN system is shown in Fig.~\ref{fig:tw}. Let $S=\{M_2,M_4\}$. According to Theorem~\ref{therm:ISO}, the LPN system is not infinite-step opaque wrt $S$, since there exists a state $(\hat{X}_1,\bar{X}_1)$ that $\hat{X}_1 \cap \bar{X}_1 = \{M_2\} \subseteq S$.

\hfill $\diamond$
\eexm

\textbf{Remark 1}:
We discuss the computational complexity of the construction of the modified TW-observer for the verification of infinite-step opacity. By Algorithm~\ref{alo:TW}, in the worst case, there are at most $2^{|X|} \times 2^{|X|}$ states and $|E_o| \times 2^{|X|} \times 2^{|X|}+|E_o| \times 2^{|X|}$ transitions in the modified TW-observer. Therefore, the complexity of the proposed algorithm is of ${\cal O}(|E_o| \times 2^{|X|} \times 2^{|X|})$. In \cite{yin2017new}, Yin and Lafortune claim that there are $|E_o| \times 2^{|X|} \times 2^{|X|}$ transitions in the TW-observer, but actually there are $2 \times |E_o| \times 2^{|X|} \times 2^{|X|}$ transitions since they just concurrent composition the two observers and the mark of the transitions on the two observers is different. Therefore, our algorithm is more efficient than that in \cite{yin2017new}.

\subsection{Verification of the K-step opacity}

In this subsection, we use the K-reduced TW-observer, which was proposed in \cite{yin2017new}, to check K-step opacity.
We denote as ${\cal B}^k_{tw}=(Q_k,E,\allowbreak f^k_{tw},q_{k0})$ the K-reduced TW-observer of the BRG. The K-reduced TW-observer is constructed to search all the states that can be reached from the initial state by observations whose length of the second elment is smaller than or equal to K.
The K-reduced TW-observer of a BRG can be constructed by applying Algorithm 1 in \cite{yin2017new}, and Theorem 7 in \cite{yin2017new} can be directly applied on BRG.

\begin{figure}
  \centering
  \includegraphics[width=0.32\textwidth]{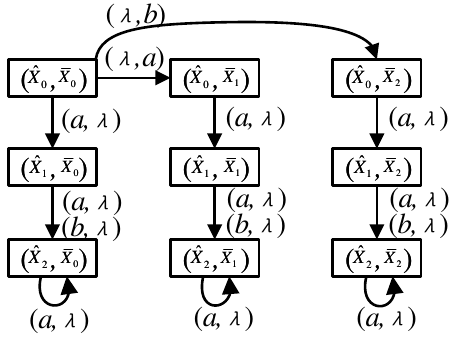}\\
  \caption{The K-reduced TW-observer of the LPN system in Fig.~\ref{fig:LPN}.}\label{fig:ktw}
\end{figure}

\them\label{therm:KSO}
Let $G$ be an LPN system, ${\cal B}^k_{tw}=(Q_k,E,\allowbreak f^k_{tw},q_{k0})$ be the K-reduced TW-observer of its BRG, and $S$ be a secret. System $G$ is K-step opacity with respect to $S$ if and only if $\forall q_k=(q_k(1),q_k(2))\in Q_k$, such that
$$q_k(1)\cap q_k(2) \nsubseteq S \vee q_k(1)\cap q_k(2) = \emptyset.$$
\ethem

In simple words, an LPN system is K-step opaque with respect to $S$ if and only if for any state $q_k$ in the K-reduced TW-observer such that the intersection of the first and second elements of $q_k$ does not belong to the secret or is empty.

\exm\label{eg:KSO}
Consider again the LPN system in Fig.~\ref{fig:LPN} whose $T_u$-induced subnet is acyclic, where the secret $S=\{M_2,M_4\}$.
Let $K=1$, thus from the initial state $(\hat{X}_0,\bar{X}_0)$, only states $(\hat{X}_0,\bar{X}_1)$ and $(\hat{X}_0,\bar{X}_2)$ are 1 step away from the second element of the initial state. And then from the three states, we search the other states though first element of these states.
Therefore, the K-reduced TW-observer of the LPN system is shown in Fig.~\ref{fig:ktw}.
According to Theorem~\ref{therm:KSO}, the LPN system is not K-step opaque wrt $S$, since there exists a state $(\hat{X}_1,\bar{X}_1)$ that $\hat{X}_1 \cap \bar{X}_1 = \{M_2\} \subseteq S$.
\hfill $\diamond$
\eexm

\section{Conclusion}\label{sec:con}

In this paper, infinite-step opacity and K-step opacity of labeled Petri nets are proposed and approaches to verify them are provided.
Under an acceptable assumption on the secret, we proved that the infinite-step opacity and K-step opacity can be checked by the basis reachability graph (BRG) and its two-way observer (TW-observer). Thus, infinite-step opacity and K-step opacity can be verified using BRG analysis rather than reachability graph analysis, which provides advantages in terms of computational complexity. And we also show that the modified TW-observer can be effectively applied to reduce the computational complexity of the solution. For Petri nets whose unobservable subnet is acyclic, the two opacity properties can be decided by constructing the TW-observer of the BRG.

Our future research will continue to focus on the computational complexity of these two opacity properties, and try to find new methods to analyze in a more efficient way.

%
%
%
%
%
%
%
%
%
%
%

\section*{Acknowledgment}

This work has been partially supported by the National Natural Science Foundation of China under Grant No. 61803317 and Grant No. 61950410604,
the Fundamental Research Funds for the Central Universities under Grant No. 2682018CX24, the Sichuan Provincial S\&T Innovation Project under Grant No. 2018027, and the Key Program for International S\&T Cooperation of Sichuan Province under Grant No. 2019YFH0097.
It has also been partially supported by Project RASSR05871 MOSIMA financed by Region Sardinia, FSC 2014-2020, annuity 2017, Subject area 3, Action Line 3.1.

\ifCLASSOPTIONcaptionsoff
  \newpage
\fi

\bibliographystyle{IEEEtran}
\bibliography{infinite-step-opacity}

\end{document}